\documentstyle[12pt,graphicx,amssymb]{article}
\let\section=\subsection     \let\subsection=\subsubsection                

\newcommand{\pdd}{P.~Danielewicz}

\newcommand{\prc}[1]{Phys.\ Rev.\ C {\bf #1}}

\newcommand{\be}{\begin{equation}}
\newcommand{\ee}{\end{equation}}
\newcommand{\ba}{\begin{eqnarray}}
\newcommand{\ea}{\end{eqnarray}}
\newcommand{\bfig}{\begin{figure}}
\newcommand{\efig}{\end{figure}}

\newcommand{\etal}{{\em et al.}}                

\begin{document}
\begin{center}
   {\large \bf NUCLEAR PHASE TRANSITIONS}\\[2mm]
   {\large \bf IN TRANSPORT THEORY}\\[5mm]
{PAWE\L}  DANIELEWICZ\\[5mm]
{\small \it National Superconducting Cyclotron Laboratory and\\
Department of Physics and Astronomy, Michigan State University,
\\
East Lansing, Michigan 48824, USA\\[8mm]}
\end{center}

\begin{abstract}\noindent
Questions concerning nuclear phase transitions are addressed
within the transport theory.  The~issue of thermal equilibrium
in the region of the liquid-gas phase transition is
investigated.  Signatures of the transition to the quark-gluon
plasma are looked for.
\end{abstract}

\section{Introduction}
Under the idealized conditions of thermal equilibrium,
the~liquid-gas phase-transition for nuclear matter is
a~necessity~\cite{dan79}.  However, is there enough time in the
reactions to reach the equilibrium?  The~numerical QCD lattice
calculations exhibit a~phase transition to the quark-gluon
plasma (QGP).  However, can the excited dense phase leave
behind any clear signals in reactions?

We shall try to address the above questions within the
transport model based on the Boltzmann equation.

\section{Boltzmann Equation}
The Boltzmann equation has the same general form
relativistically as nonrelativistically:
\be
{\partial f \over \partial t} + {\partial \epsilon_{\bf p} \over
\partial {\bf p} }
\, {\partial f \over \partial {\bf r}} -
{\partial \epsilon_{\bf p} \over
\partial {\bf r} } \, {\partial f \over \partial {\bf p}}
  =  {\cal K}^{\rm in} \, (1 - f) - {\cal K}^{\rm out} \, f
\, .
\ee
Here, $\epsilon_{\bf p}$ is the single-particle energy,
$\partial
\epsilon_{\bf
p} / \partial {\bf p}$ is the velocity, $ - \partial
\epsilon_{\bf p} / \partial {\bf r}$ is the force, and ${\cal
K}^{\rm in}$ and ${\cal K}^{\rm out}$ are the feeding and
removal rates, respectively.
The~single-particle energy and momentum, $(\epsilon_{\bf p},
{\bf p})$, form a~4-vector.  Ensuring covariance in a~model can
be aided by using a~scalar potential $U_s$ dependent on the
scalar density $\rho_s = \sum_X A_X \int d{\bf p} \, {m \over
\epsilon_p} \, f_X$.  In our model, we also add to
the energies
a~noncovariant contribution accounting for the Coulomb
repulsion, isospin dependence, and finite-range effects.
The~single-particle energies are then
\be
\epsilon_{p} = \sqrt{p^2 + (m^0 + U_s)^2} + U_v \, ,
\ee
with
\be
U_s =
{-a \, {{\xi}} + b \, {{\xi }}^\nu \over 1 +
({{ \xi }}/2.5)^{\nu - 1}} \, ,
\ee
$\xi = \rho_s/\rho_0$, and
\be
U_v = V_{\rm coul} +  d \, \nabla^2 \rho   + c \, t_3 \, \rho_T
\ee
where the isospin density is $\rho_T = \sum_X
t_3^X \, \rho_X $.

Besides nucleons and mesons, we consider in our
model the lightest clusters ($A \le 3$) whose phase-space
distribution functions follow similar transport equations to
those for the other particles.
An~obvious issue with the clusters is that of their production.
Beyond
constituents, a~minimum of one additional nucleon must be
involved in the process, as a~catalyst.  Let us take the
deuteron as an example.  In the kinetic limit
of low rates, the rate for deuteron production in
three-nucleon collisions
is given by the the matrix
element for production squared convoluted with the
$\delta$-functions for the energy-momentum conservation and
product of statistical factors:
\ba
\nonumber
\cal{K}^{\rm in} ({\bf p}) &= & \sum_{N=n,p} \int
d{\bf
p}_N \, d{\bf p}_p \, d{\bf p}_n \, d{\bf p}_N'\, \overline
{|{\cal{M}}^{npN\rightarrow Nd}|^2} \,\\ \nonumber && \times
\delta({\bf
p} + {\bf p}_N - {\bf p}_p - {\bf p}_n - {\bf p}_N') \,
\\ && \times
 \delta (\epsilon_d
+\epsilon_N - \epsilon_p - \epsilon_n - \epsilon_N')  \,
f_p \, f_n \, f_N' \, (1 - f_N) \, .
\ea
This apparently just transfers the problem of production to the
issue of determining the matrix element for production.
Because of the time-reversal invariance, though, the~production
process run backwards in time becomes the breakup process, as
illustrated in Fig.~\ref{dnp}, and, correspondingly,
the two processes share the matrix element squared:
\bfig
\centerline{\includegraphics[angle=0,
width=.65\linewidth]{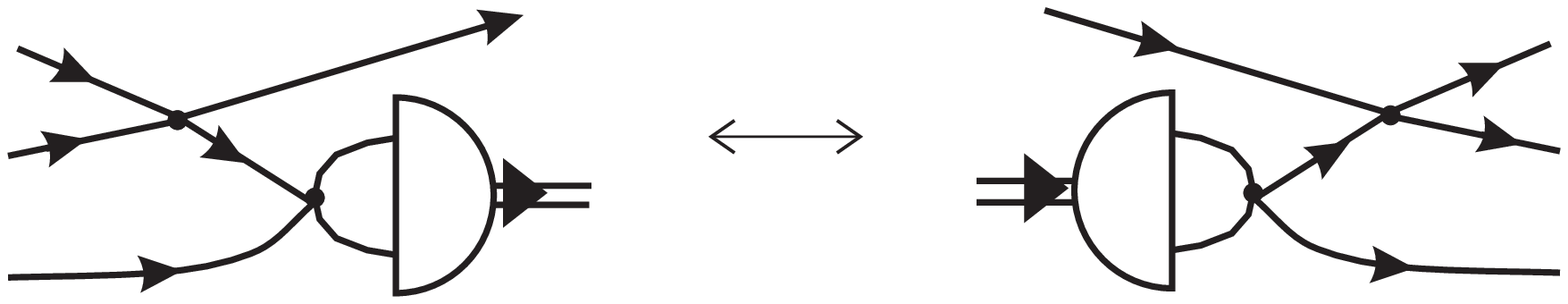}}
\caption
{The deuteron production process run backwards in time
becomes a~breakup process.}
\label{dnp}
\efig
\be
\overline{|{\cal{M}}^{npN\rightarrow Nd}|^2} =
\overline{|{\cal{M}}^{Nd\rightarrow Nnp}|^2} \, .
\ee
The~squared element for the breakup is proportional to the
breakup cross section
\be
\overline{|{\cal{M}}^{Nd\rightarrow Nnp}|^2}
\propto {d \sigma_{Nd\rightarrow Nnp} \over d{\bf p} \,
d\Omega} \, ,
\ee
and, thus, the~data on breakup can be used to describe
production.  The~production of $A=3$ clusters can be handled in
a~similar manner~\cite{dan92}.

\section{Isospin Asymmetry in Low-Density Nuclear Matter}

Under equilibrium, in the phase-transition region for
isospin-asymmetric nuclear, the~asymmetry is decreased
in the
denser liquid phase.  This is both because the interactions
play a~greater role in the liquid than in the gas phase, and
they favor the symmetry, and because of the Pauli principle.
The~enhanced symmetry leaves a~pronounced asymmetry in the gas
phase.  This was recently discussed by M\"{u}ller and
Serot~\cite{mul95}, and earlier by Barranco and
Buchler~\cite{bar80}, and by Glendenning~\cite{gle92}, and is
illustrated in Fig.~\ref{phase}.
\bfig
\centerline{\includegraphics[angle=0,
width=.77\linewidth]{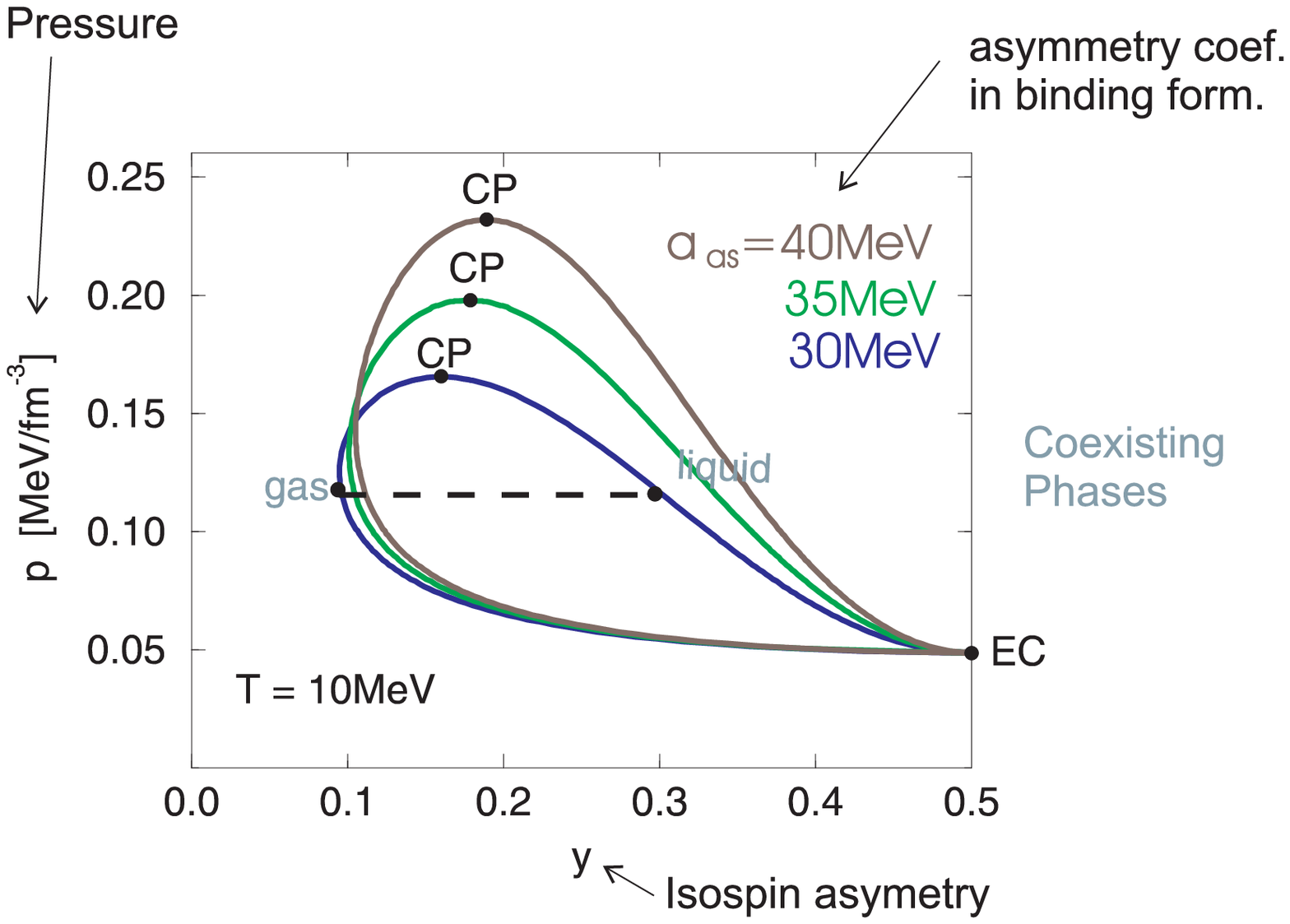}}
\caption
{Liquid-gas phase-coexistence in the plane
of pressure and isospin asymmetry at $T=10$~MeV,
for different parametrizations of interactions
in nuclear matter, after~Ref.~\protect\cite{mul95}.}
\label{phase}
\efig
The~difference in asymmetry may be as large as by a~factor
of~3.  An~obvious issue is whether pronounced asymmetries can
be seen in reactions and used as a~signature for the phase
coexistence.

The problem with the gas phase is in its low density.
If~anything corresponding to that phase is emitted, it will
most likely just escape to the vacuum.  The~seemingly only
practical
way to trap the phase and reach any kind of state of phase
coexistence is in the neck region between two slow nuclei, with
the matter in the nuclei playing the role of a~liquid phase.
This has a~further advantage that the gas phase is then well
localized in the velocity space.  Dempsey~\etal~\cite{dem96}
studied relative yields of different isotopes emitted from
xenon-tin reactions at 55~MeV/nucleon and found large
enhancement in the relative abundance of neutron-rich fragments
towards midrapidity, as illustrated in~Fig.~\ref{dempsey}.
\bfig
\centerline{\includegraphics[angle=0,
width=.94\linewidth]{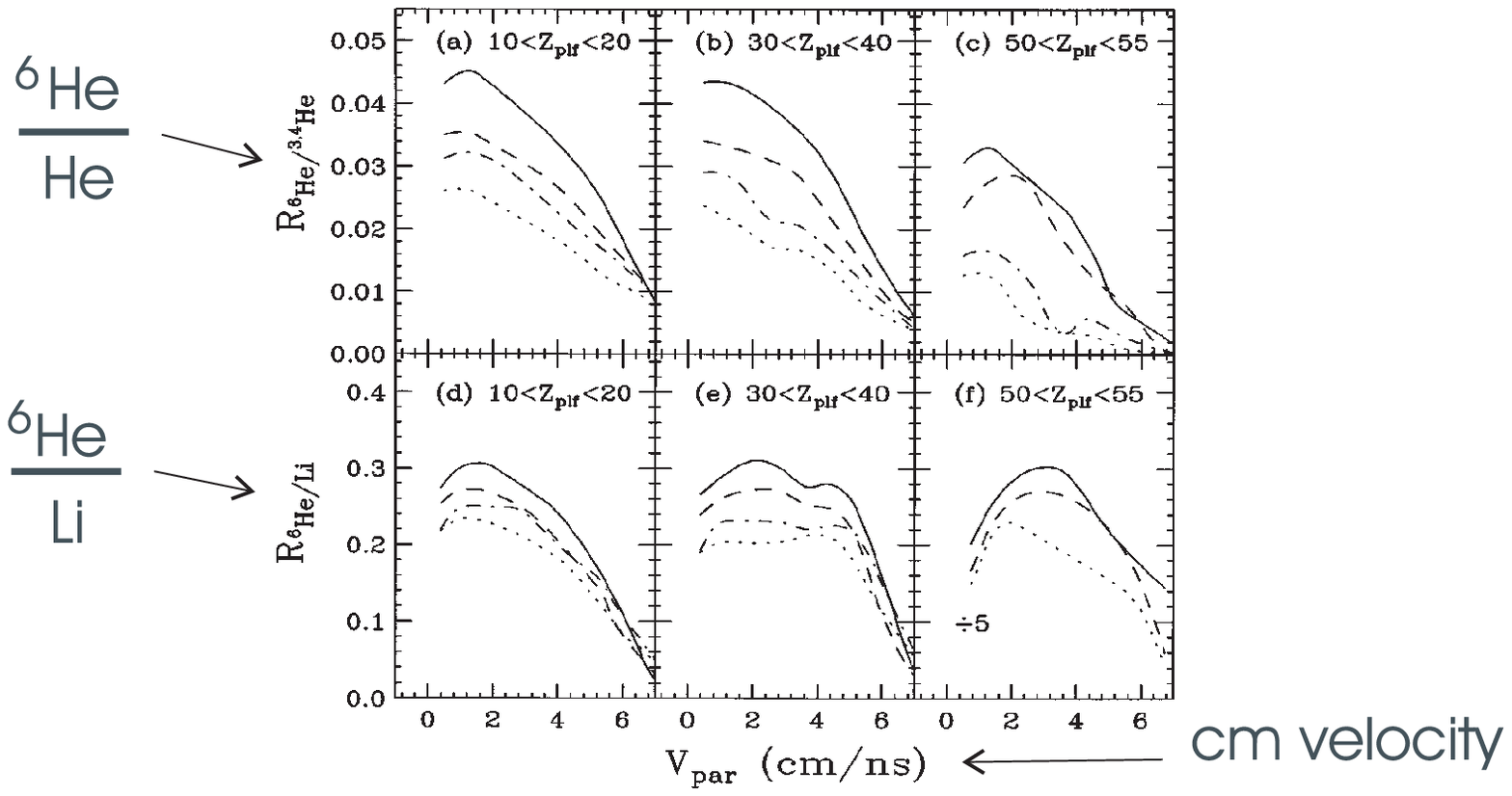}}
\caption
{Relative isotope ratios in Xe-Sn collisions
as a~function of the velocity along the beam
in~cm~\protect\cite{dem96}. Different lines represent results
from different isotopic Xe-Sn combinations.}
\label{dempsey}
\efig

To verify whether the observed enhancements might signal
effects of the phase equilibrium, calculations of the reactions
have been carried out~\cite{sob97} within the transport model
with isospin asymmetry in the interactions and in the Pauli
principle, and with the inclusion of light clusters, as
discussed before.
Figure~\ref{lee} shows the contour plots for nucleon, proton,
and
neutron density, projected onto the plane of the
$^{136}$Xe + $^{124}$Sn reaction, at different times, and
the neutron-to-proton ratio along the axis joining the two
nuclei.
\bfig
\centerline{\includegraphics[angle=0,
width=.80\linewidth]{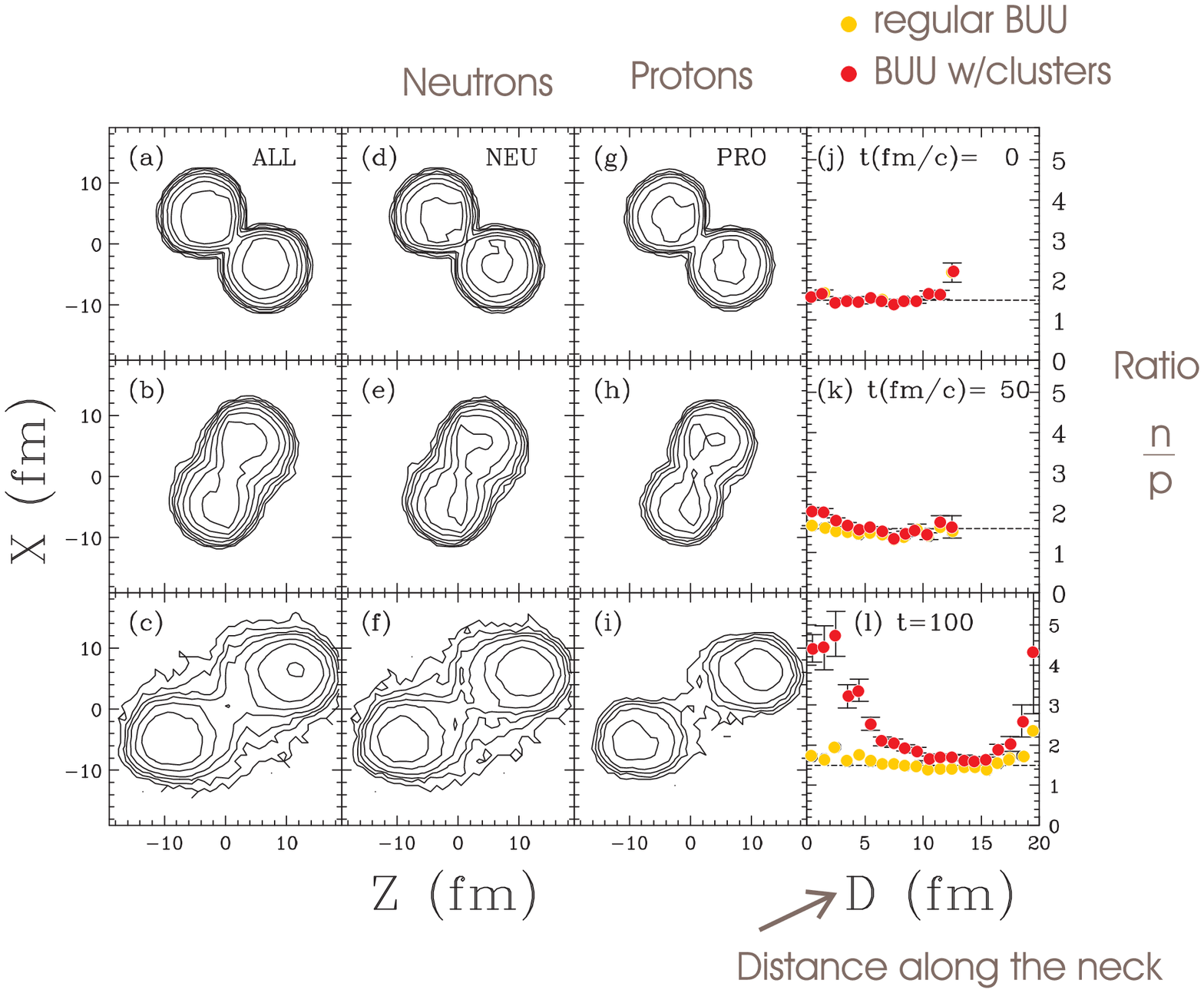}}
\caption
{
Contour plots for nucleon, proton, and neutron density
projected onto the plane of
$^{136}$Xe +
$^{124}$Sn reaction, at different times, and
the neutron-to-proton ratio along the axis joining the two
nuclei~\protect\cite{sob97}.
}
\label{lee}
\efig
The ratio is shown both including the protons and neutrons
bound in clusters and excluding them.  When the
simulation is carried without clusters, the~results for the
ratio are similar to
those for {\em all} protons and neutrons.  It is seen that the
ratio
for all nucleons just slightly exceeds the N/Z ratio of~1.5 for
the whole system.  When the clusters are excluded, the ratio
reaches high values, up to~4.

The~reason for the enhancement in the calculation is the
deuteron formation which robs the remainder of the neck region
of protons enhancing the asymmetry for the remainder.  One can
expect that, in reality, alpha particles play an~analogous
role to deuterons
leading to the asymmetries observed experimentally.  On the
other hand, the~calculations show that the overall isospin
asymmetry between the gas and liquid phases, due to the
favoring of the symmetry in the liquid and expected in the
equilibrium, has no time to develop.  This brings in the
general issue of equilibration within the neck region.

The~isospin ratio may be viewed as a~convenient variable to
assess the equilibration within the phase transition region
that, seemingly is not reached.  Calculationally, it is far
easier to determine the isospin asymmetry than
e.g.~pressure or temperature.  On~the other hand, one can argue
that the isospin equilibrates in a~slow diffusive process
(as e.g.\ smell, aided, though, in the air by convection) while
such quantities as pressure or overall density equilibrate
rapidly in a mechanical process mediated pressure waves.
Thus, isospin could not be a~reliable measure.

Let us take $L$ as a~distance over which adjustments of
a~thermodynamic quantity are to propagate and $t$ as time
that the propagation takes.
In a~diffusive process, a~particle undergoing $N$ collisions
covers a typical distance:
\be
L^2 = N \, \lambda^2 \, ,
\ee
where $\lambda$ is the mean-free-path.  The~collision number is
related to the net time by $t = N \, \lambda /v$, where $v$ is
a typical particle speed.  The~characteristic time for the
isospin adjustements is then:
\be
t_{\rm iso} = {L^2 \over \lambda \, v} \, .
\ee
The~time for the density and pressure adjustements, on the
other hand, is
\be
t_{\rm mech} = {L \over c_s} \simeq {L \over v} \, .
\ee
The~ratio of the two times,
\be
{t_{\rm iso} \over t_{\rm mech}} \simeq {L \over \lambda}
\ee
would have been large, if $\lambda$ were short.  However, in
the neck region
$\rho \sim \rho_0/3$ and thus $\lambda \sim 6$~fm.  At the same
time the distance from the nuclei to the center of the neck is
$L \sim 5$~fm and we are in the Knudsen limit!  There is no
difference between the times for isospin and density
adjustements, both are of the order of~30~fm.  Even for
a~Knudsen gas,
the~equilibration could take place
through a~contact with a~thermostat, i.e.\ nuclei, but that
does not happen in the simulation.  We~find that we are {\em
not} dealing with a~phase transition near equilibrium.

\section{Model for the Transition to Quark-Gluon
Plasma}

Let us now turn to the other important phase transition in
nuclear physics, to QGP.
The~phase
transition is approached when hadrons increase in number,
pushing out from their region the nonperturbative vacuum.
This can make the hadrons lighter as the condensates
responsible for hadronic masses are removed.
The~transition is expected to occur when hadrons completely
overlap.  The~lattice calculations for baryonless matter point
to a~phase transition at the temperature of $T_c \sim 170$~MeV
corresponding to the pion density $\rho_{\pi} \simeq 0.5
\left({T_c \over \hbar} \right)^3 \sim 2 \, \rho_0$.  If the
phase transition is associated with the hadrons filling up all
space, the~transition should be expected also in the relatively
cold matter when hadron density is raised to comparable values
due to compression, rather than thermal production.  The idea
is
further illustrated with a~simple quasiparticle
model~\cite{gos98}.

\begin{figure}
\centerline{\includegraphics[angle=0,
width=.61\linewidth]{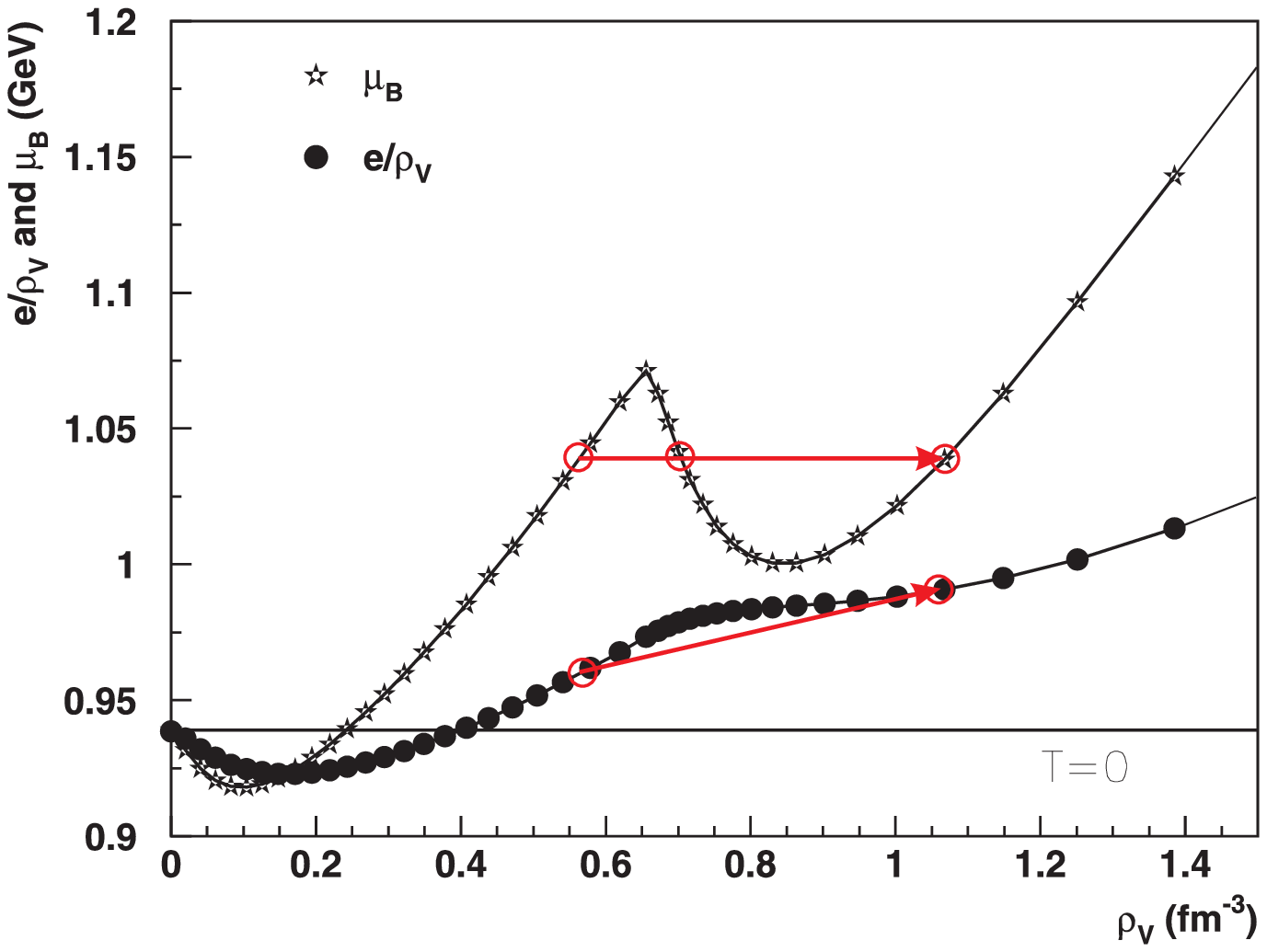}}
\caption{Chemical potential and energy per baryon
at $T=0$ as a~function of baryon density.
The~high-density phase transition is indicated with arrows.}
\label{T0}
\end{figure}

The~model~\cite{gos98}, based on the hadronic degrees of freedom,
produces a~decrease of masses in the vicinity of the phase
transition, a~large drop in masses across the transition,
and
a~dramatic increase in the number of the degrees of freedom in
the transition.  At the same time, the properties of
ground-state nuclear matter are correctly described.  The~model
is specified by giving the energy density as a~function of
particle phase distributions in the form of kinetic part and
interaction corrections depending on two densities, scalar and
vector:
\be
e = \sum_X \int d{\bf p} \, \epsilon_{\bf p}^X \,
 f^X ({\bf p})
+ e_s(\rho_s) + e_v(\rho ) \, ,
\label{e=}
\ee
where
\be
\rho_s = \sum_X \int d {\bf p} \, {m^X \, m_0^X
\over \sqrt{p^2 + m^{X2} }} \, f^X({\bf p}) \, ,
\ee
and
$
\rho = \sum_X A^X \int d {\bf p} \, f^X({\bf p}) $.
The scalar density counts all hadrons while the vector density
is just  the baryon density.  As hadron density increases,
the~hadron masses drop, from~(\ref{e=})
\be
\epsilon_{\bf p}^X = {\delta e \over \delta f^X ({\bf p})}
=
\sqrt{ p^2 + \left(S(\rho_s) \, m_0^X \right)^2} +
A^X \, U(\rho) \, ,
\ee
by a~common (for simplicity) factor~$S$ in our model.  In
a~thermally excited system the lowering of the masses leads to
the
production of more hadrons.  Instability can occur
leading to a~phase transition across
which the masses drop and the number of the degrees of freedom
rapidly increases.  We adopt simple power parametrizations of
the mass modification factor and of the vector potential to get
consistency with lattice calculations and ground-state
nuclear-matter properties,
\be
S = \left(1 - 0.54 \, (\mbox{fm}^3/\mbox{GeV}) \, \rho_s
\right)^2 \, ,
\ee
and
\be
U = {a \, (\rho/\rho_{0})^2 \over 1 + b \, {\rho/\rho_{0}
} + c \, (\rho/\rho_{0})^{5/3}}
\ee
where
$a=146.32$~MeV, $b=0.4733$, $c=51.48$.  With the~transition in
the baryonless limit at $T_c = 170$~MeV,
we find a~corresponding transition at $T=0$,
taking the matter from the baryon density of $\rho \approx 3.5
\, \rho_0$ to $7 \, \rho_0$, cf.\ Fig.~\ref{T0}.

\section{Elliptic-Flow Excitation Function}

A~possible way of detecting the high-energy nuclear phase
transition is through the associated changes in the speed of
sound.  Thus, e.g.\ in the phase-transition
region in Fig.~\ref{T0}, the speed of sound in the long
wavelength limit ($c_s = \sqrt{\partial p/ \partial e}$)
vanishes.

A~sensitive measure of the speed of sound or pressure compared
to the energy density early on in the reactions is the elliptic
flow.  The~elliptic flow is the anisotropy of transverse
emission at midrapidity.  At AGS energies, the~elliptic flow
results from a~strong competition~\cite{sor97} between
squeeze-out
and in-plane flow.
In~the early stages of the collision, the spectator nucleons
block the path of participant
hadrons emitted toward the reaction plane;  therefore the
nuclear matter is
initially squeezed out preferentially orthogonal to the
reaction plane.
In the later stages of the reaction,
the~geometry of the participant region (i.e.~a~larger
surface area exposed in the direction of the reaction plane)
favors in-plane
preferential emission.

The squeeze-out contribution to the elliptic flow and the
resulting net
direction of the flow depend on two factors: (i)~the pressure
built
up in the compression stage compared to the energy density, and
(ii)~the passage time
for removal of the spectator shadowing.
In~the hydrodynamic limit, the~characteristic time for the
development of expansion
perpendicular to the reaction plane is $\sim R/c_s$, where
$R$~is the nuclear
radius. The~passage time is
$\sim 2R /(\gamma_0
\, v_0)$, where $v_0$ is the c.m.\ spectator velocity.
The~squeeze-out
contribution should then reflect the ratio
\begin{equation}
{c_s \over \gamma_0
\, v_0 } \, .
\label{ratio}
\end{equation}

According to~(\ref{ratio}) the squeeze-out contribution should
drop with the increase in energy, because of the rise in~$v_0$
and then in~$\gamma_0$.  A~stiffer equation of state (EOS)
should yield a~higher
squeeze-out contribution.  A~rapid change in the stiffness
with baryon density and/or excitation energy should be
reflected in a~rapid change in the elliptic flow excitation
function.  A~convenient measure of the elliptic flow is the
Fourier coefficient
$\langle
\cos{2 \phi} \rangle \equiv v_2$, where $\phi$ is the azimuthal
angle of a~baryon at midrapidity, relative to the reaction
plane.  When squeeze-out dominates, the~Fourier coefficient is
negative.

\begin{figure}
\centerline{\includegraphics[angle=0,
width=.76\linewidth]{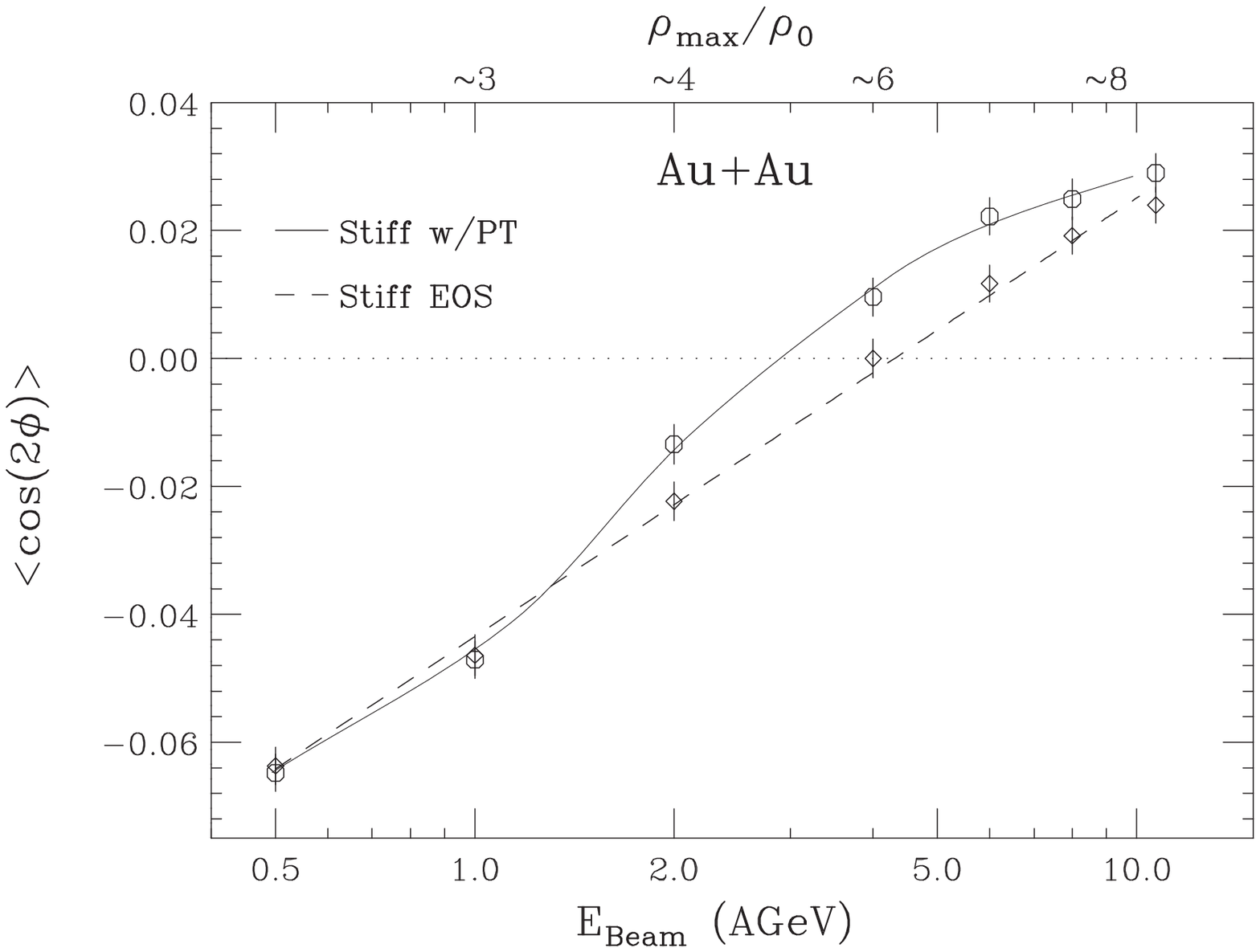}}
\caption{
      Calculated elliptic flow excitation functions for Au +
Au. The~open diamonds represent results obtained with a~stiff
EOS. The~open
circles represent results obtained with a~stiff EOS  and with
a~phase transition.
}
\label{fig4}
\end{figure}

To verify whether the expectations regarding the elliptic-flow
excitation function are realistic, we have carried out Au + Au
reaction simulations~\cite{dan98}, in the energy range of
(0.5--11)~GeV/nucleon.
The~excitation functions calculated using
a stiff EOS with a~phase transition (open circles) and
a~stiff EOS with a~smooth density dependence
are compared in Fig.~\ref{fig4}.
For low beam energies ($\lesssim \, 1$~AGeV),
the~elliptic flow
excitation functions are essentially identical because the two
EOS are either
identical or not very different at the densities and
temperatures that are
reached.  For $2 \lesssim E_{Beam} \lesssim 9$~AGeV
the~excitation function
shows  larger in-plane elliptic flow from the calculation which
includes the
phase transition, indicating that a~softening of the EOS has
occured for this beam energy range.
This deviation is in direct contrast to the esentially
logarithmic beam energy
dependence obtained (for the same energy range) from the
calculations which
assume a~stiff EOS without the phase transition.
Present data on elliptic flow from EOS, E895,
and E877 Collaborations~\cite{lac99} point to a~variation
in the stiffness of EOS in the region of
$\sim$(2--3)~GeV/nucleon, corresponding to baryon densities of
$\sim 4 \, \rho_0$.

\section{Conclusions}

To summarize, the observed large enhancements in the yields of
neutron-rich clusters in the neck region of reactions seem to
be due to the deuteron and alpha production.  Phase transition
to QGP at high $T$ in baryonless matter implies
a~transitional behavior at $T=0$ and high baryon
density.  Elliptic flow measurements can decide about
the presence or absence of the QGP transition with the
rising density.\\

        This work was supported in part
by the
National Science Foundation under Grant No.\ PHY-96-05207.

\end{document}